%% file: main.tex
\documentclass{article}
\usepackage{spconf}
\usepackage{amsmath,graphicx,hyperref}

\usepackage[utf8]{inputenc}
\usepackage{xcolor}
\usepackage{url}
\usepackage{graphicx}
\usepackage{amsmath}
\usepackage{subcaption}
\usepackage{rotating}
\usepackage{enumitem}
\usepackage{amssymb}
\usepackage{multirow}
\usepackage{booktabs}
\usepackage[normalem]{ulem} 

\usepackage{etoolbox}

\AtBeginEnvironment{equation}{\small}
\AtBeginEnvironment{equation*}{\small}
\AtBeginEnvironment{align}{\small}
\AtBeginEnvironment{align*}{\small}
\AtBeginEnvironment{gather}{\small}
\AtBeginEnvironment{gather*}{\small}

\title{Multichannel Audio Quality Assessment: Extending Pretrained Perceptual Models to Spatial Audio}

\name{Gouthaman KV, Shiv Gehlot, Vishnu Raj, Lars Villemoes, Arijit Biswas}
\address{Dolby Laboratories}

\begin{document}
\ninept

    \maketitle

    \begin{abstract}
       Accurate perceptual quality assessment is essential for evaluating and optimizing spatial audio, where perceived quality depends on both signal fidelity and inter-channel spatial relationships. However, subjective evaluation is costly, while existing perceptual models are often trained for limited channel configurations and cannot be directly applied to higher-channel-count audio. This raises the question: how can pretrained perceptual knowledge be effectively reused for multichannel spatial audio? Using 5.1-channel audio, we study four levels of multichannel integration: signal, prediction, latent, and feature and propose two learned approaches: latent-level aggregation of spatial-group representations and the feature-level Feature-Band Group Attention (FGAtt), which adaptively fuses spatial groups at the feature level before perceptual processing.
       Across five 5.1-channel test sets, FGAtt achieves the strongest overall performance, demonstrating the effectiveness of feature-level adaptation for reusing pretrained perceptual knowledge.

    \end{abstract}


\input{intro_v2.tex}
    \input{method_v2.tex}

    \input{results_v2.tex}
    \input{conclusion.tex}

    {
        \small 
        \bibliographystyle{IEEEbib}
 	    \bibliography{99_library}
    }

\end{document}

%% file: intro_v2.tex
\vspace{-3mm}
\section{Introduction}

Objective perceptual audio quality assessment is essential to audio coding, enhancement, restoration, and spatial audio systems. Although subjective tests such as MUSHRA~\cite{itu2001mushra} directly measure perceived quality, their cost and limited scalability motivate objective metrics. Conventional metrics, including PEAQ~\cite{thiede2000peaq} and ViSQOL~\cite{visqol}, use predefined perceptual representations and signal-comparison procedures~\cite{itu1387,hines2020visqol}, whereas learning-based approaches such as InSE-NET~\cite{insenet} and the Generative Machine Listener (GML)~\cite{gmlv1,gmlv2} learn representations from subjective quality data and agree more closely with human judgments than conventional signal-level metrics. Thus, learned perceptual representations offer a powerful basis for objective quality prediction.

The growing adoption of multichannel and spatial audio challenges perceptual models designed for fixed channel configurations. For instance, GMLv2~\cite{gmlv2}, a state-of-the-art learning-based perceptual model, supports mono, stereo, and binaural audio but cannot be directly applied to higher-channel formats such as 5.1 and 7.1.4. A straightforward solution is to reduce multichannel signals to a mono or two-channel representation through downmixing or binauralization~\cite{choi2008objective,kampf2010standardization,seo2013perceptual,delgado2023design}. However, such signal-level transformation can lose perceptually relevant channel-specific information and inter-channel relationships. Another approach is prediction-level late fusion, where individual channels are processed independently and their predictions are aggregated. While this avoids explicit signal reduction, it cannot model inter-channel interactions and requires additional computation for each channel. Alternatively, a standalone multichannel model can learn these interactions directly, but requires huge amount of format-specific subjective data and training and does not directly leverage perceptual knowledge from existing models such as GMLv2. As spatial formats proliferate, training separate models becomes increasingly expensive and repeatedly relearns perceptual information, motivating a scalable approach that reuses pretrained perceptual knowledge while incorporating higher-channel audio.

Consequently, this work examines how a mono/two-channel perceptual model can be adapted to higher-channel formats while preserving its perceptual knowledge. We use GMLv2 as the backbone, which achieves strong quality-prediction performance for mono, stereo, and binaural audio~\cite{gmlv2}, and study its extension to 5.1 audio. Specifically, we compare four ways of incorporating multichannel information into its processing pipeline: \emph{signal-level} transformation through downmixing or
binauralization, \emph{prediction-level} aggregation across channels
or spatial groups, \emph{latent-level} aggregation of spatial-group
representations after perceptual processing, and \emph{feature-level}
fusion before perceptual processing. Within this framework, we propose
two learned approaches that retain the pretrained GMLv2 backbone:
latent-level aggregation of spatial-group perceptual representations
and \emph{Feature-Band Group Attention} (FGAtt), which adaptively fuses
multichannel Gammatone features using input-dependent,
feature-band-specific spatial-group weights before perceptual
processing.

We evaluate the proposed strategies on five 5.1-channel test sets and
show that the feature-level adaptation method, FGAtt provides a more effective approach to reusing pretrained
perceptual knowledge than signal-domain reduction, prediction-level
aggregation, and latent-level aggregation. 
Overall, our findings highlight the importance of both the level and granularity of multichannel integration when extending pretrained mono/two-channel perceptual models to spatial audio. The results demonstrate a promising path for scaling pretrained perceptual knowledge to higher-channel-count audio while keeping the perceptual backbone frozen.

%% file: method_v2.tex
\vspace{-3mm}
\section{Methodology}
\label{sec:method}
\vspace{-2mm}
\subsection{Problem Formulation}
\vspace{-1mm}
We aim to extend GMLv2~\cite{gmlv2}, a reference-based perceptual model trained on a large collection of subjective evaluations of coded mono, stereo, and binaural audio. GMLv2 achieves stronger correlation with human judgments than conventional metrics such as ViSQOL and PEAQ, but supports inputs of at most two channels. For left and right channels $L$ and $R$, GMLv2 forms a four-component representation
$[
        L,\,
        R,\,
        (L+R)/2,\,
        (L-R)/2
    ]$. Gammatone features are extracted from these components and concatenated along the component dimension for each reference--degraded pair. A neural network then predicts the parameters of a Beta distribution, which are mapped to the MUSHRA scale to obtain the quality score. 
To extend GMLv2 to 5.1-channel audio, let $\mathbf{x}^{r}$ and $\mathbf{x}^{d}$ denote the reference and degraded multichannel audio signals, respectively. We define a multichannel integration operator $A_{\theta}^{(l)}$ at one of four levels, $l\in\{\mathrm{signal},\mathrm{prediction},\mathrm{latent},\mathrm{feature}\}$, and formulate the quality prediction as:
\vspace{-2mm}
\begin{equation}
\hat{q}
=
\mathcal{F}_{\phi}
\left(
\mathbf{x}^{r},\mathbf{x}^{d};
A_{\theta}^{(l)}
\right),
\vspace{-2mm}
\end{equation}
where $\mathcal{F}_{\phi}$ denotes the pretrained GMLv2 backbone with parameters $\phi$ kept frozen, $\hat{q}$ is the predicted perceptual quality score, and $A_{\theta}^{(l)}$ denotes the corresponding fixed or learned multichannel integration strategy. Depending on $l$, multichannel information is incorporated before or after perceptual processing, providing a common framework for comparing the four strategies while retaining the pretrained GMLv2 backbone.

\vspace{-2mm}
\subsection{Multichannel Adaptation}
We investigate four strategies for extending pretrained GMLv2 to 5.1-channel audio, categorized by the level at which multichannel information is integrated: signal, prediction, latent, or feature. Signal- and prediction-level strategies directly reuse the pretrained model, whereas latent- and feature-level strategies learn the integration within its processing pipeline. For all methods, we exclude LFE from the 5.1-channel input ${L,R,C,\mathrm{LFE},L_s,R_s}$ to avoid false penalties when codecs remove reference energy above the standard 120-Hz LFE cutoff, which subwoofer bass management would normally filter during playback.

At the \textbf{\emph{signal level}}, ITU downmixing~\cite{itu2003bs775} or binauralization~\cite{choi2008objective} transforms the 5.1-channel signal into a representation supported by GMLv2. While simple, these transformations use fixed signal-domain rules that may discard perceptually relevant spatial information, particularly when degradation characteristics vary across channels. Designing suitable fixed transformations also becomes increasingly difficult as the channel count grows, limiting their applicability to richer spatial formats.

At the \textbf{\emph{prediction level}}, GMLv2 processes the 5.1 channels either individually as mono inputs or in two-channel groups, consistent with its support for up to two channels. The resulting quality predictions are averaged to obtain the final score. Since predictions are combined only after independent processing, this approach cannot model inter-channel interactions or learn different contributions across inputs. Moreover, independent backbone evaluations increase computational cost with the number of channels or groups. The group-based variant uses the spatial grouping defined below.

These limitations motivate learned integration mechanisms that adapt to the input. At the \textbf{\emph{latent level}}, GMLv2 processes the groups independently, and learned attention weights combine their perceptual representations before quality prediction. At the \textbf{\emph{feature level}}, group features are fused before perceptual processing, enabling the frozen backbone to jointly process multichannel information. In this regard, we propose \emph{Feature-Band Group Attention} (FGAtt), which learns input-dependent, feature-band-specific weights over the groups. The groups and their common multichannel representation are defined next, followed by the respective latent- and feature-level integration strategies.

\noindent \textbf{Multichannel Feature Representation:}
The group-level prediction-, latent-, and feature-level approaches share a common spatial grouping consisting of the front pair, the surround pair, and the center channel: $\mathcal{G}=\{\mathrm{LR},\mathrm{LsRs},\mathrm{C}\}$. 
For each paired group $(x_1,x_2)$, we construct the four-component representation expected by GMLv2 $
    \left[
        x_1,
        x_2,
        (x_1+x_2)/2, 
        (x_1-x_2)/2 
    \right].$
The single center signal is duplicated according to the backbone's mono-input procedure~\cite{gmlv2}. 
Independently extracting Gammatone features from the four components yields
$\mathbf{X}^{(g)}\in\mathbb{R}^{4\times F\times D}$, $g \in \mathcal{G}$, where $F=32$ is the number of Gammatone bands and $D$ is the temporal dimension. Thus, each spatial group is mapped to the pretrained backbone's input space while preserving its spatial identity.


\noindent\textbf{Learned Latent-Level Aggregation:}
The first learned strategy introduced in this work fuses the spatial groups immediately before the GMLv2 prediction head. Each group $g\in \mathcal{G}$ is processed independently by the pretrained backbone, producing
$\mathbf{z}^{(g)}\in\mathbb{R}^{D_z}$. Each latent representation encodes quality-relevant information from its reference--degraded pair using the perceptual representation learned by GMLv2.
To learn the  relative contribution of each group, we project its latent representation and compute an attention score:
\begin{align}
    \mathbf{h}^{(g)}
    &=\tanh\left(
        \mathbf{W}_{p}\mathbf{z}^{(g)}+\mathbf{b}_{p}
    \right), \
    e_g
    &=\mathbf{h}^{(g)\top}\mathbf{v}, \ \ 
    \mathbf{v}\in\mathbb{R}^{D_h},
\end{align}
where $\mathbf{W}_{p}$, $\mathbf{b}_{p}$, and the context vector $\mathbf{v}$ are learnable. The scores are normalized across groups and used to form a weighted combination of the latent representations:
\begin{equation}
    \alpha_g
    =
    \frac{\exp(e_g)}
    {\sum_{g'}\exp(e_{g'})},
    \qquad
    \mathbf{z}_{\mathrm{agg}}
    =
    \sum_g \alpha_g\mathbf{z}^{(g)}.
    \vspace{-2mm}
\end{equation}
The frozen GMLv2 prediction head then maps $\mathbf{z}_{\mathrm{agg}}$ to the final quality score.
This approach preserves GMLv2's perceptual processing within each spatial group while learning how their representations should be combined, providing a learned alternative to prediction-level aggregation. However, it requires a separate backbone inference for every spatial group, causing computational cost to grow with the number of groups.

\noindent\textbf{Feature-Band Group Attention (FGAtt):}
Latent-level aggregation combines spatial groups only after independent perceptual processing, preventing the backbone from modeling cross-group interactions.
To enable joint processing, we instead fuse the groups at the feature level and pass the resulting representation through the frozen GMLv2 backbone. Since the perceptual contribution of each group may vary with both the input and Gammatone band, fixed fusion weights are insufficient. We therefore propose \emph{Feature-Band Group Attention} (FGAtt), which learns input-dependent, band-specific weights over spatial groups.

FGAtt follows three principles. First, temporal information is pooled only for attention estimation, leaving the temporally resolved features intact for fusion. Second, group importance is inferred jointly from the four GML-compatible components, preserving the spatial relationships encoded by the pretrained backbone. Third, a shared nonlinear mapping and cross-group normalization provide attention weights that adapt to both the input and the Gammatone feature band.

Let $g\in\mathcal{G}$ denote a spatial group with $\mathbf{X}^{(g)}\in\mathbb{R}^{4\times F\times D}$ representing its Gammatone features. We first pool the temporal dimension:
\vspace{-2mm}
\begin{equation}
    \mathbf{S}^{(g)}
    =
    \frac{1}{D}\sum_{d=1}^{D}\mathbf{X}^{(g)}_{:,:,d}
    \in\mathbb{R}^{4\times F},
    \label{eqn:temp_pool}
    \vspace{-2mm}
\end{equation}
retaining the four-component and feature-band dimensions. At each band $f$,
    $\mathbf{s}^{(g)}_{f}
    =
    \mathbf{S}^{(g)}_{:,f}
    \in\mathbb{R}^{4}$
jointly describes the activity of the four components. A shared Multi-layer Perceptron (MLP) maps this vector to a scalar attention score, 
    $e^{(g)}_{f}
    =
\operatorname{MLP}\!\left(\mathbf{s}^{(g)}_{f}\right)$.
Sharing the MLP across groups and bands keeps the adaptation module lightweight while providing a common scoring function. Because the scores are computed from the input features, they are input-dependent rather than fixed. The scores are then normalized across groups at each feature band:
\begin{equation}
    \alpha^{(g)}_{f}
    =
    \frac{\exp\!\left(e^{(g)}_{f}\right)}
    {\displaystyle\sum_{g'\in\mathcal{G}}
    \exp\!\left(e^{(g')}_{f}\right)},
    \qquad
    \sum_{g\in\mathcal{G}}\alpha^{(g)}_{f}=1.
\end{equation}
The resulting weights determine the relative contribution of each spatial group at each feature band. They are applied to the original temporally resolved features:
\begin{equation}
\mathbf{X}^{\mathrm{FGAtt}}_{:,f,d}
    =
    \sum_{g\in\mathcal{G}}
    \alpha^{(g)}_{f}\mathbf{X}^{(g)}_{:,f,d}.
    \vspace{-2mm}
\end{equation}
Thus, temporal pooling \eqref{eqn:temp_pool} affects only attention estimation, while the original temporal information is preserved during fusion. The same scalar weight is applied to all four components within a group and band, adjusting the group's overall contribution while preserving its internal GML-compatible structure.

FGAtt is applied independently to the reference and degraded representations, with attention weights determined by their respective features. The fused representations are then concatenated along the component dimension and passed to a single frozen GMLv2 backbone and prediction head. 
Unlike latent-level aggregation, which requires a separate backbone inference for each spatial group, FGAtt enables joint processing of all groups in a single inference.
It therefore provides a lightweight interface between higher-channel spatial audio and a pretrained lower-channel model while preserving the expected feature structure and temporal information. Although evaluated on 5.1 audio, the same formulation can be extended to other spatial formats by defining appropriate channel groups and their corresponding group-level representations.

%% file: results_v2.tex
\vspace{-3mm}
\section{Experiments and Results}
\vspace{-2mm}

\subsection{Experimental Setup}
\textbf{Dataset:}
Collecting subjective quality data for multichannel audio is more
involved than conventional low-channel evaluation, as perceived quality
depends on both signal fidelity and spatial characteristics. As part of
this study, we undertook a dedicated data collection effort to obtain
subjective 5.1-channel quality data for training and evaluating the
proposed learned adaptation strategies.

\noindent\textit{\textbf{Training:}}
The training corpus contains approximately 26 hours of 5.1 audio and 21K samples with subjective scores from 75 MUSHRA tests conducted over several years of multichannel codec research and development. It covers multiple generations of coding technologies: Dolby Digital, HE-AAC, Dolby Digital Plus, and AC-4, across diverse bitrates and perceptual quality levels.

\noindent\textit{\textbf{Evaluation Sets:}}
We evaluate the approaches on five independent subjective listening-test
datasets (T1--T5). Three of them (T1--T3) are derived from the MPEG Surround verification
tests~\cite{mpeg_surround}, each containing ten 5.1-channel excerpts
covering ambient, jazz, modern, orchestral, and popular music. T1, T2, and T3 were evaluated by 39, 35, and 18 listeners, respectively.
T1 covers DVB-oriented coding scenarios, including Layer-2/MPEG
Surround and Dolby Pro Logic II at 256 kb/s, as well as HE-AAC-based
multichannel systems at 64 and 160 kb/s. T2 targets music distribution
applications and compares AAC-based MPEG Surround, enhanced matrix
decoding, and Dolby Pro Logic II systems operating at approximately
192 kb/s. T3 compares high-quality and low-power MPEG Surround decoding
modes.
The remaining two test sets (T4--T5) are internal listening tests constructed
from fourteen immersive 7.1.4 excerpts by extracting the corresponding
5.1 channels. T4, evaluated by 14 listeners,
includes Opus (192 kb/s), AAC (256 kb/s), HE-AAC (192 kb/s), MPEG
Surround-based systems (192 kb/s), Dolby Digital Plus (192 kb/s), and
Dolby Digital (384 kb/s). T5, evaluated by
13 listeners, includes Opus (160 kb/s), HE-AAC (128 and 160 kb/s), and
MPEG Surround-based systems (96 and 128 kb/s).
Together, the five evaluation sets comprise 428 samples and span diverse codec families, bitrate operating points, content types, and subjective quality levels, providing a broad basis for multichannel audio quality evaluation.

\noindent\textbf{Implementation:} For signal-level experiments, we use ITU downmixing~\cite{itu2003bs775} and the Binamix binauralizer~\cite{binamix}. The learned approaches, latent-level aggregation and \emph{FGAtt}, are trained on the aforementioned training set using the objective of the original GMLv2 training~\cite{gmlv2}. The GMLv2 backbone remains frozen, while the aggregation modules are optimized with AdamW~\cite{adamw} using a learning rate of $0.001$. All methods are implemented in PyTorch and trained on two NVIDIA A10 GPUs.

\vspace{-2mm}
\subsection{Results and Discussion}
Table~\ref{tab:main} compares the evaluated methods across five test sets using Pearson correlation coefficient (PCC) and Spearman correlation coefficient (SCC) between predicted and subjective quality scores. The LFE channel is excluded throughout. The GMLv2-based methods cover prediction-, signal-, latent-, and feature-level integration, while Zimtohrli~\cite{Zimtohrli}, PEAQ~\cite{thiede2000peaq}, and ViSQOL~\cite{visqol} provide reference baselines based on established perceptual metrics with mono or two-channel inputs.

Zimtohrli~\cite{Zimtohrli} achieves an overall PCC/SCC of $0.7763/0.6872$, with performance ranging from $0.7228/0.5907$ to $0.9181/0.9138$ across the five evaluation sets. As Zimtohrli is designed for monaural audio, its multichannel extension independently evaluates each non-LFE channel and averages the resulting predictions. While this provides a straightforward extension to multichannel audio, it requires a separate metric evaluation for each channel, with computational cost increasing linearly with the number of channels. Moreover, independent channel-wise predictions do not explicitly capture inter-channel relationships that contribute to perceived spatial audio quality.

For PEAQ~\cite{thiede2000peaq}, ITU Downmix + PEAQ achieves an overall PCC/SCC of $0.4667/0.5944$, while Binauralizer + PEAQ achieves $0.4051/0.5112$. Both configurations perform substantially below their corresponding ViSQOL variants: ITU Downmix + ViSQOL achieves $0.8573/0.7455$, compared with $0.8395/0.8066$ for Binauralizer + ViSQOL. While ITU downmixing yields higher PCC with ViSQOL, binauralization achieves higher SCC and consistently outperforms ITU downmixing in SCC across all five evaluation sets. These results highlight that both the underlying perceptual metric and the choice of multichannel-to-two-channel transformation substantially affect multichannel quality prediction.

\begin{table*}
\centering
\small
\caption{Perceptual quality prediction performance on five 5.1-channel test sets.  LFE is discarded. GMLv2 is frozen. Zimtohrli evaluates each channel independently, with predictions averaged across channels.}
\label{tab:main}

\scalebox{0.8}{%
\fontsize{8}{8}\selectfont
\begin{tabular}{lcccccccccccc}
\toprule
& \multicolumn{2}{c}{T1}
& \multicolumn{2}{c}{T2}
& \multicolumn{2}{c}{T3}
& \multicolumn{2}{c}{T4}
& \multicolumn{2}{c}{T5}
& \multicolumn{2}{c}{Overall} \\
\cmidrule(lr){2-3}
\cmidrule(lr){4-5}
\cmidrule(lr){6-7}
\cmidrule(lr){8-9}
\cmidrule(lr){10-11}
\cmidrule(lr){12-13}
Method
& PCC & SCC
& PCC & SCC
& PCC & SCC
& PCC & SCC
& PCC & SCC
& PCC & SCC \\
\midrule

Zimtohrli~\cite{Zimtohrli}
& 0.7228 & 0.5907 & 0.9181 & 0.9138 & 0.8210 & 0.6948
& 0.8365 & 0.7245 & 0.8255 & 0.8129 &0.7763 & 0.6872 \\

\midrule
ITU Downmix~\cite{itu2003bs775} + PEAQ~\cite{tsp_peaq}
& 0.5846 & 0.7686 & 0.7948 & 0.8753 & 0.5199 & 0.8200
& 0.3814 & 0.5376 & 0.3023 & 0.2182 & 0.4667 & 0.5944 \\
Binauralizer~\cite{binamix} + PEAQ~\cite{tsp_peaq}
& 0.5684 & 0.7483 & 0.5970 & 0.6086 & 0.5216 & 0.8092
& 0.2969 & 0.4244 & 0.2187 & 0.0953 &0.4051 & 0.5112 \\

\midrule
ITU Downmix~\cite{itu2003bs775} + ViSQOL~\cite{hines2020visqol}
& 0.8419 & 0.7428 & 0.8739 & 0.8643 & 0.8934 & 0.7480
& 0.8682 & 0.7720 & 0.8646 & 0.7125 & 0.8573 & 0.7455 \\
Binauralizer~\cite{binamix} + ViSQOL~\cite{hines2020visqol}
& 0.8477 & 0.8791 & 0.8895 & 0.9204 & 0.8578 & 0.8798
& 0.8507 & 0.7810 & 0.8378 & 0.7549 &0.8395 & 0.8066 \\

\midrule
Mono Pred-level + GMLv2 & 0.8305 & 0.8320 & 0.8743 & 0.8651 & 0.8565 & 0.9207
& 0.7942 & 0.5653 & 0.8146 & 0.4790 & 0.7464 & 0.5249
\\
Group Pred-level + GMLv2 & 0.8759 & 0.8635 & 0.8921 & 0.8663 & 0.8846 & 0.9171 & 0.8279 & 0.7959 & 0.7808 & 0.7605 & 0.8131 & 0.7731 \\

ITU Downmix + GMLv2
& 0.9247 & 0.8759 & 0.9213 & 0.8916 & 0.9488 & 0.8893
& 0.8493 & 0.7844 & 0.7976 & 0.6891 & 0.8623 & 0.7839 \\
Binauralizer + GMLv2
& 0.8630 & 0.8498 & 0.8838 & 0.8503 & 0.8596 & 0.8716
& 0.8096 & 0.7626 & 0.7725 & 0.6947 & 0.8045 & 0.7564 \\

Latent Agg + GMLv2
& 0.8578 & 0.8414 & 0.8747 & 0.8536 & 0.9086 & 0.8740
& 0.9093 & 0.7759 & 0.9350 & 0.8364 & 0.8875 & 0.7995 \\

FGAtt + GMLv2
& 0.9001 & 0.8541
& 0.9207 & 0.8880
& 0.9106 & 0.8809
& 0.9142 & 0.7889
& 0.9415 & 0.8616
& \textbf{0.9063} & \textbf{0.8303} \\
\bottomrule
\end{tabular}%
}
\vspace{-3mm}
\end{table*}

\begin{figure}
    \centering
    \small
    \includegraphics[width=0.9\linewidth]{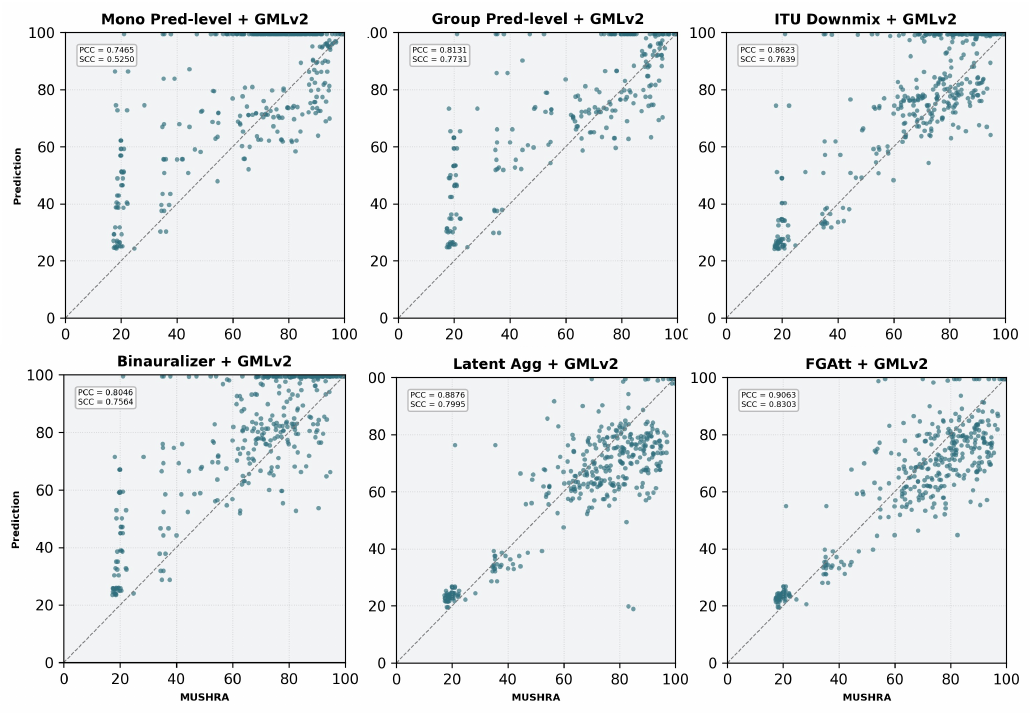}
    \caption{Overall correlation between predicted and ground-truth MUSHRA scores across GMLv2 spatial aggregation strategies. The dashed line denotes the identity function.}
    \label{fig:gmlv2_agg}
    \vspace{-6mm}
\end{figure}

Among the GMLv2-based approaches, prediction-level aggregation with
``Mono Pred-level+GMLv2" achieves an overall PCC/SCC of
$0.7464/0.5249$, while ``Group Pred-level+GMLv2" improves these
to $0.8131/0.7731$. This improvement indicates that preserving
spatially related channel structure is substantially more effective
than treating each channel independently. Both approaches, however,
require independent backbone inference for each channel or spatial group,
with predictions combined only after perceptual processing. At the signal level, ITU Downmix achieves
$0.8623/0.7839$, substantially outperforming Binauralizer at $0.8045/0.7564$. This suggests that the representation produced by ITU downmixing is more compatible with the pretrained GMLv2 representation than that produced by binauralization. However, fixed signal-domain transformations cannot adapt the contribution of different spatial groups to the input or its degradation characteristics.
Among the learned approaches, ``Latent Agg+GMLv2" further
improves the overall correlation to $0.8875/0.7995$. This improvement demonstrates the benefit of learning
spatial-group contributions rather than relying on a fixed signal-domain
transformation. However, latent aggregation still requires a separate
GMLv2 inference for each spatial group, combining the groups only after perceptual processing. The feature-level \emph{FGAtt} achieves the strongest overall
performance, with PCC/SCC of $\mathbf{0.9063/0.8303}$. 
It improves over latent aggregation in both PCC and SCC and consistently outperforms it across all five evaluation sets. Unlike latent aggregation, FGAtt combines
the spatial-group representations before perceptual processing,
allowing the frozen GMLv2 backbone to jointly process the fused
representation with a single backbone inference. Fig.~\ref{fig:gmlv2_agg} visualizes the overall correspondence between
predicted and ground-truth MUSHRA scores for the GMLv2-based
approaches. The closer alignment with the identity function is consistent with the quantitative results, with \emph{FGAtt} showing the strongest overall agreement across the evaluated samples.

The per-set results further show the complementary behavior of ITU
Downmix and FGAtt. ITU Downmix approach performs particularly
strongly on T1--T3, with PCC/SCC of $0.9247/0.8759$,
$0.9213/0.8916$, and $0.9488/0.8893$, respectively. FGAtt achieves
$0.9001/0.8541$, $0.9207/0.8880$, and $0.9106/0.8809$ on the same
sets. On T4 and T5, however, ITU Downmix drops to
$0.8493/0.7844$ and $0.7976/0.6891$, whereas FGAtt achieves
$0.9142/0.7889$ and $0.9415/0.8616$. 
Thus, while ITU Downmix performs strongly on T1--T3, FGAtt provides more consistent performance across the five evaluation sets and achieves higher correlations on T4 and T5.
The stronger performance of FGAtt on T4 and T5 is consistent with the benefit of adaptive feature-level integration when fixed signal-domain transformations may not adequately preserve the relative contributions of different spatial groups. ITU downmixing combines multiple channels before perceptual analysis, potentially reducing spatially localized information. In contrast, FGAtt adapts the contribution of each spatial group independently at each Gammatone feature band, allowing the representation presented to GMLv2 to retain different spatial contributions across the perceptual feature space. While the attention weights do not directly identify individual distortion locations, this adaptive representation provides a mechanism for the frozen backbone to jointly exploit complementary information from multiple spatial groups.

Overall, the results demonstrate that the level at which multichannel information is integrated substantially affects the effectiveness of extending pretrained GMLv2. Prediction-level and latent-level approaches preserve the backbone's perceptual processing but require independent inference for each spatial group, while signal-level approaches avoid this cost at the expense of relying on a fixed transformation. FGAtt combines the advantages of learned integration and joint processing by adaptively fusing spatial groups before perceptual processing, while keeping the GMLv2 backbone frozen.

\noindent\textbf{Ablation of Attention Granularity:}
Having established the effectiveness of feature-level adaptation, we next examine the granularity of attention in FGAtt. As described in Sec.~\ref{sec:method}, FGAtt estimates a separate attention weight for each spatial group at each Gammatone feature band, while applying the same weight to the four GML-compatible components within the group. We first remove the feature-band dimension, assigning a single weight to each spatial group across all bands. This reduces PCC/SCC from $0.9063/0.8303$ to $0.8912/0.8128$, indicating that allowing spatial-group contributions to vary across feature bands is beneficial. We then introduce component-level attention within each spatial group, allowing the four GML-compatible components to receive separate weights in addition to the band-specific group weights. This configuration achieves a PCC/SCC of $0.8803/0.8045$, lower than FGAtt. The results suggest that feature-band-dependent spatial-group weighting captures the key benefit of the proposed feature-level adaptation, while further decomposing the GML-compatible representation within each group does not provide additional benefit in this setting. Overall, the ablation supports preserving the internal GML-compatible structure while learning feature-band-dependent spatial-group attention.

%% file: conclusion.tex
\vspace{-5mm}
\section{Conclusion}
\vspace{-2mm}
We studied how a pretrained perceptual model can be extended to 5.1-channel audio by comparing signal-, prediction-, latent-, and feature-level integration. Beyond fixed signal- and prediction-level strategies, we proposed two learned methods: latent-level aggregation, which weights representations from independently processed spatial groups, and Feature-Band Group Attention (FGAtt), which adaptively fuses spatial-group features before perceptual processing. Across five 5.1-channel test sets, FGAtt achieved the strongest overall performance using a single frozen-backbone inference. These results show that the integration stage is critical and that feature-level adaptation enables effective reuse of pretrained perceptual knowledge. The framework can extend to richer formats through appropriate channel grouping and backbone-compatible representations.